\documentclass[aps,prl,amsfonts,showpacs,twocolumn]{revtex4} 
\usepackage{epsfig,amsopn}
\usepackage{graphicx}
\usepackage{amsmath}
\usepackage{natbib}

\def\be{\begin{equation}}
\def\ee{\end{equation}}
\def\bea{\begin{eqnarray}}
\def\eea{\end{eqnarray}}

\def\no{\nonumber}
\begin{document}

\title{Exact phase diagram of quasispecies model with mutation rate modifier}
\author{Apoorva Nagar}
\email{apoorva@kias.re.kr}
\altaffiliation[Present address:~]{Department of Theory and Bio-Systems, Max Planck Institute of Colloids and Interfaces,
Golm, D-14424 Potsdam, Germany}
\affiliation{ School of Physics, Korea Institute for Advanced Study, Seoul 130-722, South Korea}
\author{Kavita Jain}
\email{jain@jncasr.ac.in}
\altaffiliation[Also at~]{Evolutionary and Organismal Biology Unit}
\affiliation{ Theoretical Sciences Unit, Jawaharlal Nehru Centre
  for Advanced Scientific Research,  Jakkur P.O., Bangalore 560064,
  India}

\date{\today}

\begin{abstract}
We consider an infinite asexual population with a mutator allele which
can elevate mutation rates. With probability
$f$, a transition from nonmutator to mutator state occurs but the
reverse transition is forbidden. We find that at $f=0$, the population
is in the state with  
minimum mutation rate and at $f=f_c$, a phase transition occurs between a
mixed phase with both nonmutators and mutators and a pure mutator
phase. 
We calculate the critical probability $f_c$ and the total mutator
fraction $Q$ in the mixed phase exactly. Our predictions for $Q$ are
in agreement with those seen in microbial populations in static environments. 
\end{abstract}
\maketitle

In the absence of recombination, biological evolution is driven by two 
competing processes namely selection that tends to localise the population
around a fitness peak and mutation which has the opposite effect of
delocalising it. Extensive theoretical and experimental studies have
shown that there exists an error threshold  beyond which the
mutational load (fitness reduction) 
becomes too high to be compensated by selection
pressure  \cite{Eigen:1971,Jain:2007b}. For this reason, and because most
mutations are known to have deleterious effect
\cite{Sturtevant:1937,Kibota:1996,Drake:1998,Lynch:1999}, the 
spontaneous mutation rate is expected to be 
minimum subject to physicochemical constraints and physiological costs \cite{Kimura:1967,Liberman:1986}. 

However,
mutators with mutation rate higher than the wild type can be produced
when for example, an organism is unable to neutralize mutagens such
as radiation or repair DNA damage during replication 
\cite{Miller:1996}. In fact,   hypermutable strains in high frequency 
have been found in some cancerous cells \cite{Loeb:2003} and in 
natural isolates of certain pathogenic bacteria
which persist for many years inspite of the presence of
antibiotics \cite{LeClerc:1996}. 
Subpopulations with $10-100$ times higher 
mutation rates have also been seen to arise spontaneously 
 in long term adaptation experiments on {\it E. Coli} 
\cite{Gross:1981,Trobner:1984,Mao:1997,Sniegowski:1997,Boe:2000}. 
Due to the
persistence of mutators at long times, it is important to study their
role when mutation-selection
balance has been reached.

\begin{figure}
\includegraphics[width=0.7 \linewidth,angle=0]{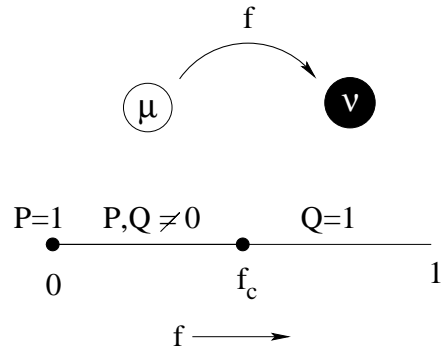}
\caption{Schematic phase diagram of the quasispecies model with
  mutation rate modifier. With probability $f$, the nonmutators (mutation probability $\mu$) change to mutators (mutation probability $\nu$). The pure nonmutator phase occurs when $f=0$ and pure mutator phase for $f \geq f_c$. The system is in the  mixed phase for $0<f<f_c$.}
\label{fphase}
\end{figure}

Here we consider the mutator problem within the framework of 
quasispecies model which is defined on the genotypic space and assumes an
infinite population \cite{Jain:2007b}. The transition from a nonmutator to mutator state occurs with probability $f$ but the reverse reaction is  ignored as it has a much smaller probability  than $f$ \cite{Drake:1998}. 
  While the mutators are thus continually generated, they are selected
against due to high mutational load. At sufficiently high $f$, we may expect the mutators to take  over the whole population. However  the possibility of such a transition has not been considered in previous studies on smooth fitness landscapes \cite{Taddei:1997,Kessler:1998,Tenaillon:1999,Johnson:1999b}.    
Here we show that the system can be in one of the following three
phases (see Fig.~\ref{fphase}): a phase with only nonmutators ($f=0$), a mixed phase in which both mutators and
nonmutators are present ($f <
f_c$) and a phase with only mutators ($f  \geq f_c$). The critical
probability $f_c$  at which a phase transition occurs between the mixed 
phase and a pure mutator phase is found exactly. We also calculate the
total mutator fraction exactly for which approximate expressions have
been obtained in the past \cite{Tenaillon:1999,Johnson:1999b}.    

We consider a haploid asexual population of infinite size evolving in
a static environment. An individual in the population is represented by a 
binary sequence $\sigma=\{\sigma_1,...,\sigma_L\}$ with $L$ loci 
where  $\sigma_i=0$ or $1$. Each sequence is endowed 
with fitness $F(\sigma)$ which is proportional to the average number
of offspring produced per generation. As there is considerable experimental 
evidence that the individual loci can contribute independently to the
genome fitness \cite{Visser:2007}, we work with multiplicative fitness 
$F(\sigma)=\prod_{i=1}^L (1-s)^{\sigma_i}$ 
where the selection coefficient $s \in [0,1)$. Assuming that independent
point mutations occur during the replication process, a sequence
$\sigma'$  mutates to $\sigma$ with a probability given by the
mutation matrix $M_{\mu}(\sigma,
\sigma')=\mu^{d(\sigma,\sigma')}(1-\mu)^{L-d(\sigma,\sigma')}$ 
where $d(\sigma,\sigma')=\sum_{i=1}^L
\sigma_i+\sigma_i'-2 \sigma_i \sigma_i'$ is the
Hamming distance between the sequences  
$\sigma$ and $\sigma'$ and $\mu$ is the mutation probability per locus per
generation. The mutation rate modifier allele is modeled by attaching  an additional bit with each sequence which controls the mutation rate but does
not affect the fitness. A nonmutator  
sequence has a spontaneous mutation probability $\mu$ per locus 
per generation while the one  
with mutator allele mutates with probability $\nu=\lambda \mu$ where 
$\lambda \geq 1$.  With forward 
probability $f$, the mutator allele is obtained from the nonmutator
and the reverse reaction is neglected. 

Thus the average fraction $P(\sigma,t)$ and $Q(\sigma,t)$ of the nonmutator and the mutator respectively at generation $t$ evolves
deterministically according to the 
following {\it coupled} nonlinear difference equations:
\bea
P(\sigma,t+1) &=& \frac{(1-f) \sum_{\sigma'} M_\mu(\sigma, \sigma') F(\sigma')
  P(\sigma',t)}{W(t)} {}  \label{Pk}\\
Q(\sigma,t+1) &=& \frac{\sum_{\sigma'} M_\nu(\sigma, \sigma') F(\sigma')
  Q(\sigma',t)}{W(t)} {} \no \\
&+&\frac{f \sum_{\sigma'} M_\mu(\sigma,\sigma') F(\sigma')
  P(\sigma',t)}{W(t)} ~.\label{Qk}
\eea
The average fitness $W(t)
=\sum_{\sigma} F(\sigma) \left[P(\sigma,t)+
  Q(\sigma,t) \right]$ in the denominator of the above equations 
ensures the normalisation condition $\sum_{\sigma}
P(\sigma,t)+Q(\sigma,t)=1$ is satisfied. We are interested in the
steady state  
when these frequencies become time-independent. The steady state phase
diagram is summarised in Fig.~\ref{fphase}. 
If $f=0$, although the steady state fractions $P(\sigma)$ and
$Q(\sigma)$ obey similar equations, the minimum mutation rate is
chosen and the population is in a pure nonmutator phase with a quasispecies localised around the fitness peak as there is no error threshold for multiplicative fitness \cite{Jain:2007b}. For $f \neq
0$, while  the nonmutator population  reduces due to nonzero forward
rate, they are 
favored over the mutators since the latter have the tendency to delocalise due to elevated mutation rates. 
Due to this
competition, we may anticipate a phase transition in the $\lambda-f$
plane between the mixed
and pure mutator phase. Note that this transition is different from the error threshold transition in which the population delocalises from the fitness peak beyond a critical error rate  \cite{Jain:2007b}. 

Before discussing the phase transition, we first demonstrate 
that for generic initial conditions, the population is in the state
with minimum mutation rate when $f=0$. Writing $P(\sigma,t)=
Y(\sigma,t)/X(t)$  and $Q(\sigma,t)= Z(\sigma,t)/X(t)$ 
in Eqs.~(\ref{Pk}) and (\ref{Qk})  where $X(t)=\sum_{\sigma'}
Y(\sigma',t)+Z(\sigma',t)$ and 
\be
X(t+1)=\sum_{\sigma} F(\sigma) \left[Y(\sigma,t)+Z(\sigma,t) \right]~,
\label{Xeqn}
\ee 
we find that the unnormalised variables
$Y(\sigma,t)$ and $Z(\sigma,t)$ obey linear uncoupled equations
given by \cite{Jain:2007b} 
\bea
Y(\sigma,t+1)=\sum_{\sigma'} M_\mu(\sigma, \sigma') F(\sigma')
Y(\sigma',t) \label{Yunnorm} \\
Z(\sigma,t+1)=\sum_{\sigma'} M_\nu(\sigma, \sigma') F(\sigma')
Z(\sigma',t)~.
\label{Zunnorm}
\eea
The solution of the above equations is of 
the following product form,
\be
Y(\sigma,t)=\prod_{i=1}^L y_0^{1-\sigma_i} y_1^{\sigma_i}~,~Z(\sigma,t)=\prod_{i=1}^L z_0^{1-\sigma_i} z_1^{\sigma_i}
\label{YZprod}
\ee
where the time-dependent fractions $y_k$ and $z_k~,~k=0,1$ obey
Eqs.~(\ref{Yunnorm}) and (\ref{Zunnorm}) for
$L=1$. This can be verified, for example, for $Y(\sigma,t)$ by  using
the above ansatz in 
Eq.~(\ref{Yunnorm}) whose right hand side (RHS)  can be 
expressed as a product over $L$ terms, 
\bea
&&\sum_{\sigma'} \prod_{i=1}^L  (1-\mu) (1-s)^{\sigma'_i} y_0^{1-\sigma'_i} y_1^{\sigma'_i} \left(\frac{\mu}{1-\mu}\right)^{\sigma_i+\sigma'_i-2 \sigma_i \sigma'_i} \no\\
&=&\prod_{i=1}^L \mu^{\sigma_i} (1-\mu)^{1-\sigma_i} y_0 (t)+
\mu^{1-\sigma_i} (1-\mu)^{\sigma_i} (1-s) y_1(t) \no \\
&=&\prod_{i=1}^L y_0^{1-\sigma_i}(t+1) ~y_1^{\sigma_i}(t+1) \no
\eea
where we have used the evolution equations for $y_0, y_1$  to arrive
at the desired result. 
For an initial condition 
in which all the population is at the 
least fit sequence $\sigma^{(0)}=\{1,1,...,1\}$ with 
unnormalised nonmutator population $\alpha \neq 0$ and mutator
population $\beta$,   
the one locus fractions $y_k$ and $z_k$ can be straightforwardly computed and we find that 
the ratio $z_0(t)/y_0(t) \sim \left( \frac{\beta}{\alpha} \right)^{1/L}
\frac{\kappa_-^t(\nu) -\kappa_+^t(\nu)}{\kappa_-^t(\mu)- \kappa_+^t(\mu)}$
where
\be
\kappa_{\pm}=\frac{(2-s) (1-\mu)\pm \sqrt{4 \mu^2 (1-s)+s^2
    (1-\mu)^2}}{2}~.
\label{kappa}
\ee
Since $\kappa_- < \kappa_+$ and $\kappa_+(\nu) < \kappa_+(\mu)$, it follows that $z_0/y_0$  
vanishes when $t \to \infty$. Using this  in the expression for the average fitness $W(t)$, it follows that   the steady state fitness $W=\kappa_+^L(\mu)$ as in the pure nonmutator phase.  

\begin{figure}
\includegraphics[width=0.6 \linewidth,angle=270]{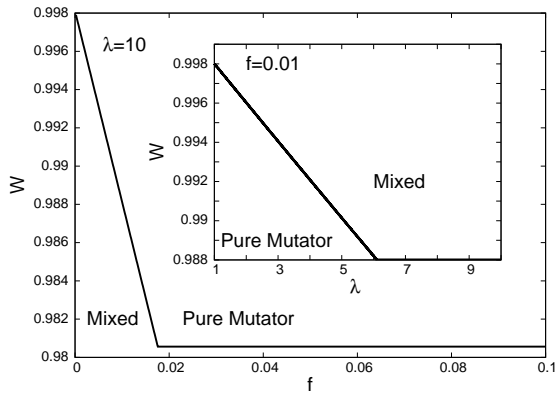}
\caption{Average fitness $W$  as a function of  $f$ (main)
  and $\lambda$ (inset) for $L=20, \mu=10^{-4}$ and $s=0.05$.}
\label{Wvsflam}
\end{figure}
\begin{figure}
\includegraphics[width=0.6 \linewidth,angle=270]{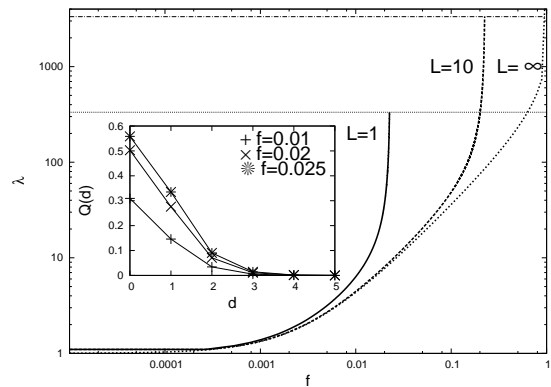}
\caption{Phase diagram in the $f-\lambda$ plane obtained using
  Eq.~(\ref{phasedia}) with $\mu L=0.003$ and $s=0.05$. The horizontal
  lines show the maximum possible value of $\lambda$ as 
  $\nu=\lambda \mu< 1$. Inset: Mutator distribution $Q(d)$ as a function of the distance $d$ from the master sequence for $L=10$ and $f_c=0.025$.}
\label{flamphase}
\end{figure}

We now consider the interesting situation when the forward rate $f$ is
nonzero. In the following, we will first find the average fitness
$W_<$  for $f < f_c$ and $W_>$ for $f > f_c$ and then
determine the critical point $f_c$ by matching $W_<$ and $W_>$ at the 
transition. The steady state equations for the population fractions, unlike the
time-dependent ones, can not be 
linearised by passing to $Y$ and $Z$ variables due to
Eq.~(\ref{Xeqn}). As a consequence, due to the normalisation factor
$W$ in the denominator, Eq.~(\ref{Pk}) for the nonmutator fraction is
coupled to the mutator fraction. 
However on summing over $\sigma$ on both sides of  
Eq.~(\ref{Pk}) in the steady state, we find that provided the
nonmutator fraction  
$P=\sum_\sigma P(\sigma)$ is nonzero, the average fitness
$W$ does not depend on the mutator fraction and writeable as
\be
W=\frac{(1-f) \sum_\sigma F(\sigma) P(\sigma)}{\sum_\sigma P(\sigma)}~,~P \neq 0
\label{fitreln}
\ee
thus leading to an uncoupled nonlinear equation for
$P(\sigma)$. Eliminating $W$ from Eq.~(\ref{Pk}) using the above
equation and writing ${\tilde P}(\sigma)=P(\sigma)/\sum_{\sigma'}
P(\sigma')$, we see that ${\tilde P}(\sigma)$ obeys the quasispecies 
equation for a population without mutation rate modifier. In this
case, the normalised steady state distribution is known to
be given by \cite{Woodcock:1996}, 
\be
{\tilde P}(\sigma)= \prod_{i=1}^L {\tilde p}_0^{1-\sigma_i} {\tilde p}_1^{\sigma_i}\label{prod}
\ee
where ${\tilde p}_0, \tilde{p}_1$ are the
solutions of the corresponding one locus model and the average fitness
is given by $\kappa_+^L(\mu)$ as seen in the pure nonmutator phase. 
From Eq.~(\ref{fitreln}), we thus obtain $W_<=(1-f) \sum_\sigma
F(\sigma) {\tilde P}(\sigma)=(1-f) ~ \kappa_+^L(\mu)$. 
Contrary to na{\"i}ve expectation, the fitness $W_<$ is unaffected by 
the mutation rate $\nu$. This result is 
consistent with the reduction principle that requires the average
fitness to be maximised \cite{Kimura:1967,Liberman:1986}.  This can
happen if only the nonmutators contribute to the fitness thus
minimising the mutational load due to mutators. But as the forward rate is
nonzero, this contribution is reduced by a factor $1-f$. Besides $P
\neq 0$ (mixed phase),  $P=0$  is also a solution of Eq.~(\ref{Pk}). 
For $f > f_c$ phase in which this solution is valid, the mutator
population $Q(\sigma)=\prod_{i=1}^L q_0^{1-\sigma_i} q_1^{\sigma_i}$ (see Fig.~\ref{flamphase}) 
and the average fitness  $W_>= \kappa_+^L(\nu)$.

A plot of average fitness as a function of $f$ and $\lambda$ is shown
in Fig.~\ref{Wvsflam}. With increasing $f$ as the population goes from
the mixed to pure mutator phase, the average fitness decreases to a 
constant since the nonmutator fraction vanishes for $f > f_c$. As
shown in the inset, in
the pure mutator phase, the fitness decreases with increasing 
$\lambda$ since the increase in mutation rate decreases the population
at the top of the fitness landscape. In the mixed phase, as discussed
above, the fitness is constant in $\lambda$. 
Matching the fitnesses $W_<$ and $W_>$ at the critical point,  the
phase boundary in the $f-\lambda$ plane is obtained (see
Fig.~\ref{flamphase}) ,  
\be
(1-f_c)^{1/L}
=\frac{(2-s) (1-\nu_c)+\sqrt{4 \nu_c^2 (1-s)+s^2
    (1-\nu_c)^2}}{(2-s)(1-\mu)+\sqrt{4\mu^2(1-s)+s^2(1-\mu)^2}} \label{phasedia}
\ee 
 When $\nu=\mu$ or $s=0$, the critical point $f_c=0$ and
the population is always in the pure mutator phase. 

\begin{figure}
\includegraphics[width=0.6 \linewidth,angle=270]{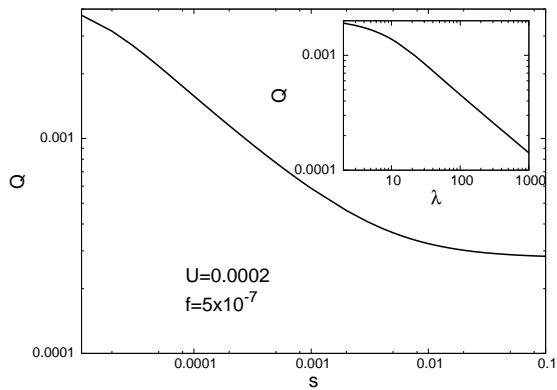}
\caption{Variation of mutator fraction $Q$ with $s$ (main) and $\lambda$
  (inset) in the mixed phase obtained using Eq.~(\ref{exactQ}) when $n \to \infty$.} 
\label{qlams}
\end{figure}

So far we have discussed the quasispecies model for arbitrary genome
length $L$. To compute the total fraction $Q=\sum_{\sigma}Q(\sigma)$
of mutators in the mixed phase, we now consider the model defined by
Eqs.~(\ref{Pk}) and 
(\ref{Qk}) when the genome length $L \to \infty$ and the mutation
probabilities $\mu, \nu \to 0$ with $U=\mu L$ and $V=\nu L$ finite. 
In this limit, a sequence at a Hamming distance $k$ from the fittest
mutates  to one at distance $j$ with mutation probability 
$M_{U}(j,k)$ which is a Poisson distribution with mean $U$ for $k \leq
j$ and zero otherwise \cite{Higgs:1994,Woodcock:1996}. Furthermore, 
the average fitness $W_<= (1-f) e^{-U}$ and $W_>=e^{-V}$
so that the critical probability $f_c=1-e^{-{V+U}}$, independent of
$s$. The  fractions $P(k)$ and $Q(k)$ of nonmutators and mutators
respectively at a Hamming distance $k$ from the fittest sequence obey
the following equations \cite{Johnson:1999b}, 
\bea
P(k) &=& \sum_{k'=0}^{k}\frac{U^{k-k'}}{(k-k')!} (1-s)^{k'} P(k') \label{Pkuni}\\
Q(k) &=& c_1  \sum_{k'=0}^{k}
\frac{V^{k-k'}}{(k-k')!} (1-s)^{k'} Q(k')+ c_2 P(k) \label{Qkuni}
\eea
where the coefficient $c_1=(1-f_c)/(1-f)$ and $c_2=f/(1-f)$. 
  Using the Eqs.~(\ref{Pkuni}) and (\ref{Qkuni}), one can write down
  the recursion relations for the  generating function $G(z)=\sum_{k=0}^{\infty} z^k Q(k)$ and 
$H(z)=\sum_{k=0}^{\infty} z^k P(k)$ 
with $G(1)=1-H(1)=Q$. On applying these recursion equations $n$ times
and setting $z=1$, we obtain
\bea
Q &=&c_1^n e^{V \sum_{k=0}^{n-1} (1-s)^k} G((1-s)^n) {} \no\\
&+&c_2 (1-Q)  \sum_{m=0}^{n-1} c_1^m e^{(V-U) \sum_{k=0}^{m-1} (1-s)^k}~.
\label{exactQ}
\eea
In the limit $n \to \infty$, the first term on the RHS drops out for $c_1 < 1$ ($f < f_c$) and  the sum can be expressed in terms of an incomplete gamma function by replacing the infinite sum by an integral. For biologically realistic situations for which  $f \ll s \ll V-U$ (see below), the sum can be calculated approximately to yield 
\be
Q \approx \frac{ f \sqrt{2 \pi}}{f \sqrt{2 \pi}+ (1-f) \sqrt{(V-U)s}}
\ee
which increases with $f$ but decreases with $U, \lambda$ and $s$ (also
see
Fig.~\ref{qlams}).  

As an application of our results, we consider a large population of bacteria  {\it E. Coli} for which the genome mutation rate 
$U \approx 2 \times10^{-4}$ \cite{Kibota:1996,Drake:1998,Boe:2000} and the spontaneous
forward transition rate $f \approx 5 \times 10^{-7}$  \cite{Ninio:1991} has been
estimated.  Our calculations show that a subpopulation of weak mutators with $\lambda=10$ can take over the
entire population if $f >  2 \times 10^{-3}$. Such a high transition rate can for example occur  in the presence of
mutagens \cite{Mao:1997}. Moreover,  our preliminary simulations on finite populations indicate that 
$f_c$   is considerably reduced due to stochastic fluctuations and 
 it  should be possible to observe this
phase transition in experiments even without mutagens.   As shown in Fig.~\ref{qlams}, we obtain $Q \sim 0.5\% -0.03 \%$ for $s \sim 10^{-5}-10^{-1}$ and $\lambda=10$ using Eq.~(\ref{exactQ}).
Such small fractions have been observed in several long term experiments on {\it E. Coli}: $0.25\%$ in \cite{Gross:1981}, $0.6\%$ in \cite{Trobner:1984} and more recently, $0.5\%$ in \cite{Mao:1997}. 

To summarise, we have presented several exact results for a
quasispecies model with mutation rate modifier. The model discussed above can well describe large populations in a stable environment and harboring less than $1\%$ of mutator subpopulation . However 
if the population is exposed to a continuously changing environment  as in infectious diseases \cite{LeClerc:1996}, the mutator fraction can be as high as $50-70\%$ and we need to consider the evolution on dynamic fitness landscapes \cite{Trobner:1984, Tenaillon:1999}.

AN acknowledges the kind hospitality of JNCASR during his visits. K.J. thanks KITP, Santa Barbara for hospitality where she learnt about
the mutator problem.


\end{document}